\documentclass[epjST]{svjour}
\usepackage{graphics}
\usepackage{amsmath}
\usepackage{amssymb}
\begin{document}
\title{Kinetic description of vacuum $e^+ e^-$ production in strong electric fields of arbitrary polarization}
%\subtitle{Do you have a subtitle?\\ If so, write it here}
\author{I.~A.~Aleksandrov\inst{1}\fnmsep\inst{2} \and V.~V.~Dmitriev\inst{3} \and D.~G.~Sevostyanov\inst{1} \and S.~A. Smolyansky\inst{3}\fnmsep\inst{4}\fnmsep\thanks{\email{smol@sgu.ru}}}

\institute{Department of Physics, St. Petersburg State University, 199034 Saint Petersburg, Russia \and Ioffe Institute, 194021 Saint Petersburg, Russia \and Department of Physics, Saratov State University, 410026 Saratov, Russia \and Laboratory of Quantum Theory of Intense Fields, Tomsk State University, 634050 Tomsk, Russia}
\abstract{
We present a detailed analysis of the self-consistent system of kinetic equations (KEs) describing electron-positron pair production from vacuum under the action of a spatially homogeneous time-dependent electric field of arbitrary  polarization. The physical significance of all the basic functions of the kinetic theory is ascertained. It is demonstrated that the total  system  of  the  KEs consists of two coupled quasiparticle and spin subsystems with their integrals of motion. A projection method is proposed in order to obtain the KE system in two particular cases: linearly polarized external electric field and (2+1)-dimensional description of quasiparticles in graphene. We also address the energy conservation law taking into account the internal plasma field and describe an alternative rigorous derivation of the KE system.
}
\maketitle
\section{Introduction}\label{sec.1}

Vacuum creation of electron-positron plasma (EPP) in the presence of strong quasiclassical electromagnetic fields is a fundamental phenomenon predicted by quantum electrodynamics (QED)~\cite{1_sauter_1931,1_euler_heisenberg,1}. Unfortunately, there is a very limited number of exactly solvable cases~(see, e.g., Refs.~\cite{3,nikishov_jetp_1970,narozhny_1970,gav_git_prd_1996,adorno_ijmpa_2017,Cherv}), which are basically used as a reference for benchmarking methods that can be implemented numerically. Some of these methods which can be utilized in studies of a broad class of various electromagnetic backgrounds are based on nonperturbative kinetic approaches. The most general method rests on the Wigner (phase) representation~\cite{5,6} (see also Refs.~\cite{19,XLT}). An alternative technique is formulated in the framework of the quasiparticle representation~\cite{7,8,9} and can be employed in the case of a spatially homogeneous external background. This approach proved to be particularly productive in the simplest case of a linearly polarized electric field which possesses an arbitrary temporal dependence. Here one can demonstrate that this method is completely equivalent to that formulated within the Wigner representation (see Refs.~\cite{5,6}). In Ref.~\cite{10} the technique was generalized in order to consider a combination of a linearly polarized time-dependent electric field and a homogeneous magnetic background. In the case of arbitrary polarization, the method based on the quasiparticle representation was developed in different ways in Refs.~\cite{11,12,13,14}. All of these approaches lead to the well-defined system of the kinetic equations (KEs).

In the present study, we perform a comprehensive analysis of the KE system in order to gain a better understanding of its structure and the role of the spin degrees of freedom. Special focus is placed on the physical interpretation of the twelve basic functions involved in the KEs. By means of the projection method, we also establish a solid connection between the general KE system and its specific form in the case of linear polarization and the low-dimensional model of graphene. This investigation is expected to become profitable for various implementations of kinetic theory, e.g., in QCD~\cite{43} and theory of graphene~\cite{4a}, and for its further development, e.g., by incorporating the photon degrees of freedom~\cite{14,15} and taking into account radiation processes~\cite{46,45,16}.

The paper is organized as follows. In Sect.~\ref{sec.2} we present the KE system and describe the notations used. We also briefly discuss a simplified version of the KEs in the case of a linearly polarized electric field. In Sect.~\ref{sec.3} we elucidate the physical meaning of the single-particle correlation functions including anomalous averages entering the KE system. In Sect.~\ref{sec.4} it is demonstrated that there is a closed group of seven KEs describing vacuum creation of EPP and vacuum polarization effects. This subsystem can be considered separately from the other five KEs, which govern the spin degrees of freedom. We also discuss how these seven KEs are reduced in the case of a linearly polarized electric field and in graphene. In Sect.~\ref{sec.5} we draw a conclusion. In Appendix we present a detailed derivation of the KE system in the framework of the instantaneous basis approach.

Throughout the article, we assume $\hbar=c=1$. The electron charge is $e=-|e|$.

\section{System of kinetic equations in quasiparticle representation}\label{sec.2}

The quasiparticle representation is attained by diagonalizing the Hamiltonian at each time instant $t$, which can be done in the case of a spatially homogeneous electric background. We describe the external field by a vector potential in the gauge $A^0 = 0$ with an arbitrary temporal dependence,
\begin{equation}\label{2.1}
  A^{\mu}(t)=(0,A^1(t),A^2(t),A^3(t)).
\end{equation}
A nonperturbative KE system determining vacuum production of EPP was derived in this representation by means of several different techniques~\cite{11,12,13,14}. It reads
\begin{equation}\label{2.2}
  \begin{aligned}
    \dot{f} &= -2(\boldsymbol{\lambda}_1 \vec{u}), \\
   \dot{\vec{f}} &= -2[\vec{f} \boldsymbol{\lambda}_2] + 2[\vec{v}  \boldsymbol{\lambda}_1] - 2\boldsymbol{\lambda}_1 u, \\
   \dot{u} &= 2(\boldsymbol{\lambda}_1 \vec{f}) +2\omega v, \\
   \dot{\vec{u}} &= \boldsymbol{\lambda}_1 (2f-1) - 2[\vec{u} \boldsymbol{\lambda}_2] + 2\omega \vec{v}, \\
   \dot{v} &= -2\omega u, \\
   \dot{\vec{v}} &= - 2[\vec{v} \boldsymbol{\lambda}_2] - 2\omega \vec{u}.
  \end{aligned}
\end{equation}
The dot and cross products of vectors $\vec{a}$ and $\vec{b}$ are denoted by $(\vec{a} \vec{b})$ and $[\vec{a} \vec{b}]$, respectively. The twelve unknown components $f$, $\vec{f}$, $u$, $\vec{u}$, $v$, $\vec{v}$ depend on time and quasimomentum $\vec{P}=\vec{p}-e\vec{A}(t)$. These functions originate from the single-particle electron and positron correlation functions
\begin{equation}\label{2.3}
\hat{f}(\vec{p},t)\sim f_{s' s}(\vec{p},t)=\langle a^{\dagger}_{s}(\vec{p},t)\, a_{s'}(\vec{p},t)\rangle=\langle b^{\dagger}_{s}(-\vec{p},t)\, b_{s'}(-\vec{p},t)\rangle
\end{equation}
and from the following combinations of the anomalous averages:
\begin{eqnarray}\label{2.4}
  \hat{u}(\vec{p},t)\sim u_{s' s}(\vec{p},t)=\frac{1}{2}\left[\langle a^{\dagger}_{s}(\vec{p},t)\, b^{\dagger}_{s '}(-\vec{p},t)\rangle+\langle b_{s}(-\vec{p},t)\, a_{s '}(\vec{p},t)\rangle\right],\\ \label{2.5}
\hat{v}(\vec{p},t)\sim v_{s ' s }(\vec{p},t)=\frac{i}{2}\left[\langle a^{\dagger}_{s}(\vec{p},t)\, b^{\dagger}_{s '}(-\vec{p},t)\rangle-\langle b_{s}(-\vec{p},t)\, a_{s '}(\vec{p},t)\rangle\right] .
\end{eqnarray}
These matrices are rewritten in the Pauli (spin) representation $(\hat{\cal{O}}=\hat{f},\hat{u},\hat{v})$:
%
%\begin{eqnarray}\label{2.6}
%  \hat{\cal{O}} &=& {\cal{O}} \mathbb{I}+(\boldsymbol{\cal{O}} \boldsymbol{\sigma}),\\ \label{2.7}
% {\cal{O}} &=& \frac{1}{2}\, \mathrm{Sp} \, \hat{\cal{O}},  \quad \boldsymbol{\cal{O}}=\frac{1}{2}\, \mathrm{Sp} \, \hat{\cal{O}} \boldsymbol{\sigma}.
%  \end{eqnarray}
%
\begin{equation}\label{2.6}
  \hat{\cal{O}} = {\cal{O}} \mathbb{I}+(\boldsymbol{\cal{O}} \boldsymbol{\sigma}),\quad {\cal{O}} = \frac{1}{2}\, \mathrm{Sp} \, \hat{\cal{O}},  \quad \boldsymbol{\cal{O}}=\frac{1}{2}\, \mathrm{Sp} \, \hat{\cal{O}} \boldsymbol{\sigma}.
\end{equation}
The indices $s$, $s' = 1$, $2$ determine the spin state, $\mathbb{I}$ denotes the identity matrix, $\sigma^k$ ($k=1$, $2$, $3$) are the Pauli matrices. %The averaging procedure in Eqs.~(\ref{2.3})--(\ref{2.5}) is carried out with respect to the {\it in} vacuum state.

The function $f(\vec{p},t)$ determines the momentum distribution of quasiparticles, whereas $\vec{f}(\vec{p},t)$ describes the spin effects. The other functions, $u$, $v$, $\vec{u}$, and $\vec{v}$, incorporate vacuum polarization effects.
%in the quasiparticle and spin subsystems. 
The physical meaning of the components involved in the system~(\ref{2.2}) will be discussed in Sect.~\ref{sec.3}. The functions $\boldsymbol{\lambda}_1$ and $\boldsymbol{\lambda}_2$ are defined as %\cite{11}
\begin{equation}\label{2.8}
  \begin{aligned}
(\boldsymbol{\lambda}_1)_i = \frac{\dot{\omega}P_i}{2\omega\omega_+}-\frac{e}{2\omega}E_i &= \frac{e}{2\omega}\left[ \frac{P_i P_k}{\omega\omega_+} -\delta_{ik} \right] E_k \equiv \frac{e}{2\omega} l_{ik} E_k,\\
(\boldsymbol{\lambda}_2)_i &= \frac{e}{2\omega \omega_+}[\vec{P} \vec{E}]_i,
  \end{aligned}
\end{equation}
where $\omega \equiv \omega(\vec{p},t)=\sqrt{m^2+\vec{P}^2}$ is quasienergy, $\omega_+ = \omega + m$, $\vec{E}(t)=-\dot{\vec{A}}(t)$ is the field strength, and $i = 1$, $2$, $3$. We also imply here and below summation over the repeated indices. Note that $(\boldsymbol{\lambda}_1 \boldsymbol{\lambda}_2)=0$.

The KE system~(\ref{2.2}) is considered together with the zero initial conditions, i.e., at $t = t_\text{in}$ we set $f = u = v = \vec{f} = \vec{u} = \vec{v} = 0$, which corresponds to the initial vacuum state. If the electric field vanishes only asymptotically, one should imply $t_\text{in} \to -\infty$.

The system (\ref{2.2}) can be derived within several independent approaches leading to the quasiparticle representation. In Ref.~\cite{11} it was obtained with the aid of the Foldy-Wouthuysen transformation. In Refs.~\cite{12,13,14} the system (\ref{2.2}) was derived by means of a transition to the nonstationary bispinor basis using the substitution $p^i \rightarrow P^i$. To this end, one can employ the following orthonormal and complete set of bispinors:
\begin{equation}\label{2.9}
  \begin{aligned}
  u^\dagger_1(\vec{p},t) &= B(\vec{p})[\omega_+,0,P^3,P_-],\\
  u^\dagger_2(\vec{p},t) &= B(\vec{p})[0,\omega_+,P_+,-P^3],
  \end{aligned}
  \qquad
    \begin{aligned}
  v^\dagger_1(-\vec{p},t) &= B(\vec{p})[-P^3,-P_-,\omega_+,0],\\
  v^\dagger_2(-\vec{p},t) &= B(\vec{p})[-P_+,P^3,0,\omega_+],
  \end{aligned}
\end{equation}
where $P_{\pm}=P^1 \pm iP^2$ and $B(\vec{p})=(2\omega\omega_+)^{-1/2}$. These bispinors satisfy~\cite{20}
\begin{equation}\label{2.10}
  \begin{aligned}
  u^\dagger_s(\vec{p},t) v_{s'}(-\vec{p},t) &= 0,\\
  u^\dagger_s(\vec{p},t) u_{s'}(\vec{p},t) &= v^\dagger_s(-\vec{p},t) v_{s'}(-\vec{p},t) = \delta_{s s'},\\
  \bar{u}_s(\vec{p},t) u_{s'}(\vec{p},t) &= \frac{m}{\omega(\vec{p},t)}\delta_{s s'},\\
  \bar{v}_s(\vec{p},t) v_{s'}(\vec{p},t) &= -\frac{m}{\omega(\vec{p},t)}\delta_{s s'},
  \end{aligned}
\end{equation}
where $\bar{w} \equiv w^\dagger \gamma^0$. An equivalent derivation of the KEs (\ref{2.2}) in the framework of  an instantaneous basis approach is presented in detail in Appendix.

The field operator in the quasiparticle representation has the following form:
\begin{equation}\label{2.psi}
\psi (\vec{x}, t)=\int \! \frac{d^3p}{(2\pi)^{3/2}} \, \mathrm{e}^{i\vec{p} \vec{x}} \, \sum_s \big [a_s(\vec{p},t) u_s(\vec{p},t) + b^\dagger_s(-\vec{p},t) v_s(-\vec{p},t) \big ].
\end{equation}
Using the bispinors~(\ref{2.9}), one can recast the standard Hamiltonian of QED into a diagonal form within the quasiparticle representation:
\begin{equation}\label{2.11}
  \hat{H}_{\rm q}(t)=\int \! d^3p \ \omega(\vec{p},t) \sum_s \big [a^\dagger_s(\vec{p},t)a_s(\vec{p},t) - b_s(-\vec{p},t)b^\dagger_s(-\vec{p},t)\big ].
\end{equation}
We also point out that the bispinor basis~(\ref{2.9}) is useful for generalization of the kinetic approach by taking into account the interaction between EPP and the quantized part of the electromagnetic field, which allows one to address other phenomena such as vacuum photoeffects, photon radiation {\it etc}. (see Refs.~\cite{15,16}).

In the case of a linearly polarized electric field
\begin{equation}\label{2.12}
  A^{\mu}(t)=(0,0,0,A^3(t)),
\end{equation}
the KE system (\ref{2.2}) turns to the well-known system of the following form~\cite{5,7,8,9}:
\begin{equation}\label{2.13}
  \begin{aligned}
  \dot{f}_{\rm lin} &= \frac{1}{2}\lambda U_{\rm lin},\\
  \dot{U}_{\rm lin} &= \lambda (1-2f_{\rm lin})-2\omega V_{\rm lin},\\
  \dot{V}_{\rm lin} &= 2\omega U_{\rm lin},
  \end{aligned}
\end{equation}
where $\lambda$ is the amplitude of vacuum transitions which reads
\begin{equation}\label{2.14}
\lambda = \frac{e E \varepsilon_\perp}{\omega^2},
\end{equation}
and $\varepsilon_\perp=\sqrt{m^2+p^2_{\perp}}$ ($p_\perp$ is the transversal momentum component).

%The KE system (\ref{2.13}) possesses an integral of motion
%
%\begin{equation}\label{2.15}
%  \begin{aligned}
%(1-2f_{\rm lin})^2+U^2_{\rm lin}+V^2_{\rm lin}=1,
%  \end{aligned}
%\end{equation}
%
%compatible with the zero initial condition.

In what follows, we will thoroughly analyze the KE system~(\ref{2.2}) starting with identifying the physical significance of all of the twelve basic components appearing in the KEs.

\section{Physical interpretation of the basic functions} \label{sec:phys_int}\label{sec.3}

To illuminate the physical meaning of the scalar functions $f$, $u$, and $v$ and the vector functions $\vec{f}$, $\vec{u}$, and $\vec{v}$, we aim to express various macroscopic mean values in terms of these components using the definitions~(\ref{2.3})--(\ref{2.6}). We are interested in quantities which can be generally represented as $K(t)=\langle \hat{K}(t) \rangle$, where
%it can, by analogy with the kinetic theory in the Wigner representation~\cite{19}, to write in terms of these functions the standard set of the integral macroscopic averaging physical values of QED $\Pi(t)=\langle \hat{\Pi}(t) \rangle$, where
%
\begin{equation}\label{3.1}
  \begin{aligned}
\hat{K}(t)=\int \! d^3x \, \hat{\kappa}(x).
  \end{aligned}
\end{equation}
We use here a shorthand notation $x \equiv (\vec{x}, t)$. The operator $\hat{\kappa}(x)$ will be chosen in the form of the operators of current density, energy-momentum tensor density, and spin tensor density. We start with the following general definitions:
\begin{eqnarray}\label{3.2}
\hat{j}^{\mu}(x) &=& e:\!\bar{\psi}(x)\, \gamma^{\mu}\psi(x)\!: \,,\\
\label{3.3}
\hat{t}^{\mu\nu}(x) &=& \frac{i}{4}:\![\bar{\psi}(x)\, \gamma^{\mu}D^{\nu}\psi(x)-(D^{*\nu}\bar{\psi}(x))\, \gamma^{\mu}\psi(x)]+(\mu\leftrightarrow\nu)\!: \,,\\
\label{3.4}
\hat{s}^{\lambda (\mu\nu)}(x) &=& \frac{1}{4}:\! \bar{\psi}(x)\, \gamma^{\lambda}\sigma^{\mu \nu}\psi(x)\!:+  \frac{1}{4}:\! \bar{\psi}(x)\,\sigma^{\mu \nu}\gamma^{\lambda}\psi(x)\!: \,,
\end{eqnarray}
where $D^{\mu}=\partial^{\mu}+ieA^{\mu}$, $\sigma^{\mu \nu}=i(\gamma^\mu \gamma^\nu -\gamma^\nu \gamma^\mu)/2$, and colons denote normal ordering. Here we employ the relevant decomposition of the field operator given in Eq.~(\ref{2.psi}).

Using the anticommutation relations, the total current, which is the integral of Eq.~(\ref{3.2}) over $\vec{x}$, can be represented as a sum of two parts: conduction current and polarization current which have the form
\begin{eqnarray} \nonumber
\hat{J}_{\mu}^{\rm cond}(t) &=& e \int \! d^3p \sum_{s,s'} \bigg \{  \bar{u}_s (\vec{p},t)\gamma_{\mu} u_{s'}(\vec{p},t) \bigg [ a^\dagger_s(\vec{p},t)a_{s'}(\vec{p},t)-\frac{\delta_{ss'}}{2} \bigg ] \\
{}&+& \bar{v}_s (-\vec{p},t)\gamma_{\mu} v_{s'}(-\vec{p},t) \bigg [ b_s(-\vec{p},t)b^\dagger_{s'}(-\vec{p},t)-\frac{\delta_{ss'}}{2} \bigg ] \bigg \}, \label{3.6}\\
\nonumber
\hat{J}_{\mu}^{\rm pol}(t) &=& e \int \! d^3p \sum_{s,s'} \Big [  \bar{u}_s(\vec{p},t) \gamma_{\mu} v_{s'}(-\vec{p},t) a^\dagger_s(\vec{p},t)b^\dagger_{s'}(-\vec{p},t) \\
{}&+& \bar{v}_s(-\vec{p},t) \gamma_{\mu} u_{s'}(\vec{p},t) b_s(-\vec{p},t)a_{s'}(\vec{p},t) \Big ]. \label{3.7}
\end{eqnarray}
The mean total charge vanishes as was expected due to the charge conservation:
\begin{equation}\label{3.8}
Q(t)  = \langle \hat{J}_{0}(t) \rangle =  e \int \! d^3p \sum_{s}\left( f_{ss}+g_{ss}-\delta_{ss}\right)= 2e \int \! d^3p\ (f+g-1) =0.
\end{equation}
Here the trace of the matrix $g_{s s'} (\vec{p}, t) = \langle b_{s'} (-\vec{p}, t) b_s^\dagger (-\vec{p}, t) \rangle$ yields a scalar function $g$ which is not involved in the system~(\ref{2.2}) due to the exact relation $f+g=1$ leading to $Q(t) = 0$ (see, e.g., Refs.~\cite{11,14}). The spatial components of the current can be obtained using the following properties of the spinor basis~(\ref{2.9}):
\begin{eqnarray}
&&u^\dagger_s (\vec{p},t)\, \boldsymbol{\alpha}\, u_{s'}(\vec{p},t) = \vec{v}_\text{g} \delta_{ss'}, \qquad v^\dagger_s(-\vec{p},t)\, \boldsymbol{\alpha}\, v_{s'}(-\vec{p},t) = - \vec{v}_\text{g} \delta_{ss'},\label{3.9}\\
&&u^\dagger_s (\vec{p},t)\, \boldsymbol{\alpha}\, v_{s'}(-\vec{p},t) = v^\dagger_s (-\vec{p},t)\, \boldsymbol{\alpha}\, u_{s'}(\vec{p},t)  =\boldsymbol{\sigma }_{ss'} - \frac{\vec{P}}{\omega \omega_+}(\boldsymbol{\sigma}_{ss'} \vec{P}),\label{3.10}
\end{eqnarray}
where $\alpha_k=-\gamma_0 \gamma_k$ and $\vec{v}_\text{g} = \vec{P}/\omega$ is the group velocity of quasiparticles. Thus,
\begin{eqnarray}\label{3.12}
J_{i}^{\rm cond}(t) &=& -e \int \! d^3p \sum_{s}\left( f_{ss}-g_{ss}\right) v_{\text{g} i} = -2e \int \! d^3p\ (2f-1)v_{\text{g} i},\\
J_{i}^{\rm pol}(t) &=& - e \int \! d^3p \sum_{s,s'} u^\dagger_s (\vec{p},t)\, \alpha_i \, v_{s'}(-\vec{p},t) u_{s's}  = 2e \int \! d^3p\ l_{ik}(\vec{p},t)\, u_k(\vec{p},t),\label{3.13}
\end{eqnarray}
where the tensor $l_{ik}(\vec{p},t)$ is defined in Eq.~(\ref{2.8}).

To calculate the energy-momentum tensor, one should make use of the equations of motion for the creation and annihilation operators since it is necessary to evaluate time derivatives in Eq.~(\ref{3.3}). From the Dirac equation, it follows that
\begin{eqnarray}
\dot{a}_{s'}(\vec{p},t)&=&-i\omega a_{s'}(\vec{p},t) - i\boldsymbol{\lambda}_2\boldsymbol{\sigma}_{s's} a_{s}(\vec{p},t)-\boldsymbol{\lambda}_1\boldsymbol{\sigma}_{s's} b^\dagger_{s}(-\vec{p},t), \label{3.14} \\
\dot{b}^\dagger_{s'}(-\vec{p},t) &=& i\omega b^\dagger_{s'}(-\vec{p},t) - i\boldsymbol{\lambda}_2\boldsymbol{\sigma}_{s's} b^\dagger_{s}(-\vec{p},t)+\boldsymbol{\lambda}_1\boldsymbol{\sigma}_{s's} a_{s}(\vec{p},t).\label{3.15}
\end{eqnarray}
%
%We will also employ the following relations:
%%
%\begin{eqnarray}
%u^\dagger_{s}\dot{u}_{s'}-\dot{u}^\dagger_{s}u_{s'} &=& v^\dagger_{s}\dot{v}_{s'}-\dot{v}^\dagger_{s}v_{s'} = 2i(\boldsymbol{\lambda}_2\boldsymbol{\sigma}_{ss'}),\label{3.16} \\
%u^\dagger_{s}\dot{v}_{s'}-\dot{u}^\dagger_{s}v_{s'} &=& -v^\dagger_{s}\dot{u}_{s'}+\dot{v}^\dagger_{s}u_{s'} =
%2(\boldsymbol{\lambda}_1\boldsymbol{\sigma}_{ss'}),\label{3.16_2}\\
%u^\dagger_{s}\alpha_k\dot{u}_{s'}-\dot{u}^\dagger_{s}\alpha_k u_{s'} &=& -v^\dagger_{s}\alpha_k\dot{v}_{s'}+\dot{v}^\dagger_{s}\alpha_k v_{s'} =
%2i [\boldsymbol{\sigma}_{ss'} \boldsymbol{\lambda}_1]_k, \label{3.17} \\
%u^\dagger_{s}\alpha_k\dot{v}_{s'}-\dot{u}^\dagger_{s}\alpha_k v_{s'} &=& v^\dagger_{s}\alpha_k\dot{u}_{s'}-\dot{v}^\dagger_{s}\alpha_k u_{s'} = 2i\delta_{ss'}(\boldsymbol{\lambda}_2)_k, \label{3.18}
%\end{eqnarray}
%%
%where we do not explicitly indicate the temporal and momentum dependences. Finally, we obtain the mean value of energy,
One obtains
\begin{equation}\label{3.19}
E_\text{q} (t)=\int \! d^3x \, \langle t_{00}(x) \rangle   = \int \! d^3p \sum_{s}\omega \left( f_{ss}-g_{ss}\right)=2\int \! d^3p\ \omega\left( 2f-1 \right)%,
\end{equation}
and the mean value of momentum ($k=1$, $2$, $3$),
\begin{equation}\label{3.20}
P_k(t)=\int \! d^3x \, \langle t_{0k}(x) \rangle = 0 .
\end{equation}

As can be seen from the equations of motion~(\ref{3.14}) and (\ref{3.15}), the Hamiltonian $H_{\rm q}(t)$ [see Eq.~(\ref{2.11})] does not completely describe the evolution of the quasiparticle system. The terms with $\boldsymbol{\lambda}_1$ and $\boldsymbol{\lambda}_2$ in Eqs.~(\ref{3.14}) and (\ref{3.15}) correspond to a unitary time-dependent transformation within the quasiparticle representation and are related to the Heisenberg picture, where the following Hamilton operators describe the vacuum polarization and spin effects, respectively:
\begin{eqnarray}\label{3.21}
\hat{H}_{\rm pol}(t)&=&-i\int \! d^3p \sum_{s,s'} (\boldsymbol{\lambda}_1 \boldsymbol{\sigma}_{ss'})\big[a^{\dagger}_{s}(\vec{p},t)b^{\dagger}_{s'}(-\vec{p},t) -b_{s}(-\vec{p},t)a_{s'}(\vec{p},t)\big],\\
\label{3.22}
\hat{H}_{\rm s}(t)&=&\int \! d^3p \sum_{s,s'} (\boldsymbol{\lambda}_2 \boldsymbol{\sigma}_{ss'})\big[a^{\dagger}_{s}(\vec{p},t)a_{s'}(\vec{p},t) -b_{s}(-\vec{p},t)b^{\dagger}_{s'}(-\vec{p},t)\big].
\end{eqnarray}
The total Hamiltonian has the form $\hat{H}(t)=\hat{H}_{\rm q}(t)+\hat{H}_{\rm pol}(t)+\hat{H}_{\rm s}(t)$,
%
%\begin{equation}\label{3.23}
%\hat{H}(t)=\hat{H}_{\rm q}(t)+\hat{H}_{\rm pol}(t)+\hat{H}_{\rm s}(t),
%\end{equation}
%
so the corresponding energy reads
\begin{equation}\label{3.24}
  \begin{aligned}
T_{00}^{\rm tot}(t)=\langle \hat{H}(t)\rangle= 2\int \! d^3p \big [\omega(2f-1) - (\boldsymbol{\lambda}_1\vec{v}) + (\boldsymbol{\lambda}_2\vec{f}) \big].
  \end{aligned}
\end{equation}
Due to the fact that the total Hamiltonian is not stationary, the concept of energy~(\ref{3.24}) is not well defined. Nevertheless, the commutation relation $\big [ \hat{H}_{\rm q}(t), \hat{\Pi}_{\rm q}(t) \big ]=0$
%
%\begin{equation}\label{3.25}
%\big [ \hat{H}_{\rm q}(t), \hat{\Pi}_{\rm q}(t) \big ]=0
%\end{equation}
%
is valid [here, $\hat{\Pi}_{\rm q}(t)$ is the quasiparticle part of $\hat{\Pi}(t)$].

For the description of the EPP spin, we will employ the Pauli spin operator~\cite{20,21,22}
\begin{equation}\label{3.26}
%\hat{S}_i=
\frac{1}{2}\Sigma_i =\frac{1}{2}
\begin{pmatrix}
\sigma_i & 0 \\
0 & \sigma_i
\end{pmatrix},
\end{equation}
which is in accordance with the spin decomposition~(\ref{2.6}). %We note that the operator~(\ref{3.26}) does not commute with the free-particle Hamiltonian $H_0 = (\boldsymbol{\alpha} \vec{p}) + \beta m$, so the corresponding vector is not conserved, which leads to {\it Zitterbewegung} of the Pauli spin~\cite{20,27,28,29}. There are numerous other definitions of the relativistic spin operator (see Ref.~\cite{30} and references therein).
The macroscopic spin %corresponding to the Pauli operator~(\ref{3.26}) 
is described by the Noether spin density~(\ref{3.4}) integrated over the spatial coordinates for $\lambda=0$ and $\mu$, $\nu = 1$, $2$, $3$,
\begin{equation}\label{3.26a}
\hat{S}_i(t)=\frac{1}{2}\int \! d^3x \, \psi^\dagger (x)\Sigma_i\psi(x).
\end{equation}
In the spinor basis~(\ref{2.9}), $S_i(t) = \langle \hat{S}_i(t) \rangle$ reads
\begin{equation}\label{3.27}
\vec{S}(t)=2 \int \! d^3p\left[\vec{f}-\frac{\vec{P}}{\omega\omega_+}(\vec{P}\vec{f})-\frac{\vec{P}u}{\omega}\right].
\end{equation}
The three terms correspond to the proper spin, spin-orbital and vacuum polarization contributions, respectively.

Besides the average quantities (\ref{3.12}), (\ref{3.13}), (\ref{3.19}), (\ref{3.20}), (\ref{3.24}), and (\ref{3.27}), the mean value of the stress tensor $T_{ik}(\vec{x},t)$ is of interest. It can be calculated by taking into account the following diagonal form of the correlation functions in the momentum space:
\begin{equation}\label{3.33}
  \begin{aligned}
\langle a^{\dagger}_s(\vec{p},t)a_s(\vec{q},t)\rangle =f_{ss}(\vec{p},t)\delta(\vec{p}-\vec{q}).
  \end{aligned}
\end{equation}
It brings us to the following result:
\begin{equation}\label{3.34}
 T_{ik}(t)  =\frac{1}{2e} \int \! d^3p  \big [ P_i j_k(\vec{p},t) + P_k j_i(\vec{p},t)  \big ],
\end{equation}
where $j_k(\vec{p},t)$ is the current density in the momentum space,
\begin{equation}\label{3.35}
 J_{k}(t) =\int \! d^3p \, j_k(\vec{p},t).
\end{equation}

In Table~\ref{table} we summarize which of the twelve basic functions are involved in the macroscopic quantities. As can be seen from here, the vacuum polarization function $v(\vec{p},t)$ takes part in none of the corresponding expressions. Its role will also be discussed in the next section.

\begin{table}
	%\small
	\centering
\setlength{\tabcolsep}{1.0em}
	\caption{Summary of which of the basic functions of the KE system~(\ref{2.2}) are involved in various macroscopic average quantities.} \label{table}
    \begin{center}
	\begin{tabular}{lcccccccccccc}
    \hline\noalign{\smallskip}
    %\hline
    \multicolumn{1}{c}{Macroscopic quantity} &
    $f$&
    $\vec{f}$&
    $u$&
    $\vec{u}$&
    $v$&
    $\vec{v}$
     \\
    \noalign{\smallskip}\hline\noalign{\smallskip}
    Conduction current $J^{\rm cond}_{i}(t)$&
    $+$&
    $-$&
    $-$&
    $-$&
    $-$&
    $-$
     \\
    % \hline
    Polarization current $J^{\rm pol}_{i}(t)$&
    $-$&
    $-$&
    $-$&
    $+$&
    $-$&
    $-$
     \\
    % \hline
    Quasiparticle energy $E_{\rm q}(t)$&
    $+$&
    $-$&
    $-$&
    $-$&
    $-$&
    $-$
     \\
     %\hline
    Polarization energy $E_{\rm pol}(t)$&
    $-$&
    $-$&
    $-$&
    $-$&
    $-$&
    $+$
     \\
     %\hline
    Spin energy $E_{\rm s}(t)$&
    $-$&
    $+$&
    $-$&
    $-$&
    $-$&
    $-$
     \\
     %\hline
    Momentum $P_{i}=0$&
    $-$&
    $-$&
    $-$&
    $-$&
    $-$&
    $-$
     \\
     %\hline
      Spin vector $S_i(t)$&
    $-$&
    $+$&
    $+$&
    $-$&
    $-$&
    $-$
     \\
     %\hline
       Stress tensor $T_{i k}(t)$&
    $+$&
    $-$&
    $-$&
    $+$&
    $-$&
    $-$
     \\
     \noalign{\smallskip}\hline
    % \hline
           \end{tabular}
    \end{center}
\end{table}

\section{Analysis of the KE system. Projection method} \label{sec:analysis} \label{sec.4}

Let us first consider some general properties of the KE system~(\ref{2.2}). Note that the system contains a closed subsystem governing the evolution of the distribution function $f$ and two vector functions $\vec{u}$ and $\vec{v}$:
\begin{eqnarray}\label{4.1}
\nonumber
  \dot{f} &=& -2(\boldsymbol{\lambda}_1 \vec{u}),\\
  \dot{\vec{u}} &=& \boldsymbol{\lambda}_1 (2f-1)-2[\vec{u} \boldsymbol{\lambda}_2]+2\omega \vec{v},\\\nonumber
  \dot{\vec{v}} &=& -2[\vec{v} \boldsymbol{\lambda}_2]-2\omega \vec{u}.
\end{eqnarray}
This subsystem describes the process of EPP vacuum creation in electric fields of arbitrary polarization. The rest part of the system~(\ref{2.2}) involves the spin function $\vec{f}$ and two scalar vacuum polarization functions $u$ and $v$ which correspond to the spin excitation processes:
\begin{eqnarray}\label{4.2}
\nonumber
  \dot{\vec{f}} &=& -2[\vec{f} \boldsymbol{\lambda}_2]+2[\vec{v} \boldsymbol{\lambda}_1]-2 \boldsymbol{\lambda}_1 u, \\
  \dot{u} &=& -(\boldsymbol{\lambda}_1 \vec{f}) + 2\omega v,\\\nonumber
  \dot{v} &=& -2\omega u.
\end{eqnarray}
These two subsystems are related by means of the common vector function $\vec{v}$, which can be considered as a source in the spin subsystem. The subsystem (\ref{4.1}) has the following integral of motion:
\begin{equation}\label{4.3}
\frac{1}{4}(1-2f)^2+\vec{u}^2+\vec{v}^2 = {\rm const}.
\end{equation}
The subsystem (\ref{4.2}) possesses another integral of motion:
\begin{equation}\label{4.4}
\vec{f}^2+u^2+v^2=2\int \limits^t_{t_\text{in}} \! dt' \, \vec{f}(t')[\vec{v}(t') \boldsymbol{\lambda}_1(t')] + {\rm const}.
\end{equation}
The constants in Eqs.~(\ref{4.3}) and (\ref{4.4}) are determined by the initial conditions regarding the time instant $t_\text{in}$, when the external field is getting switched on.

\subsection{Projection method} \label{sec:analysis_proj}

Our next task is to discuss a transition from the general KE system~(\ref{2.2}) to the KEs~(\ref{2.13}) for the case of a linearly polarized electric field. The starting point here is the comparison between the first equation of the subsystem (\ref{4.1}) and that of the system (\ref{2.13}). Let us introduce the unit vector $\vec{e}(\vec{p},t)$ via
\begin{equation}\label{4.12}
\boldsymbol{\lambda}_1=\vec{e}\Lambda,
\end{equation}
where, according to Eq.~(\ref{2.8}),
\begin{equation}\label{4.13}
\Lambda =|\boldsymbol{\lambda}_1|=\left[  \frac{e^2E_i E_k}{4\omega^2} \left( \delta_{ik} - \frac{P_i P_k}{\omega^2} \right) \right]^{1/2}.
\end{equation}
Let us consider the case~(\ref{2.12}) of the external field with a constant direction. According to Eq.~(\ref{2.8}), we have
\begin{equation}\label{4.14}
(\boldsymbol{\lambda}_1)_{i, \rm lin}=\frac{e}{2\omega} \, l_{i3}E(t).
\end{equation}
Now it is easy to see that
\begin{equation}\label{4.14a}
\Lambda_{\rm lin}=|(\boldsymbol{\lambda}_1)_{\rm lin}|=\frac{|\lambda|}{2},
\end{equation}
where $\lambda$ is defined by Eq.~(\ref{2.14}), so $|\Lambda_{\rm lin}| \sim |E(t)|$. In order to properly treat a possible alternating sign of the field strength $E(t)$ in accordance with Eq.~(\ref{4.14}), it is enough to redefine the unit vector $\vec{e}$ in Eq.~(\ref{4.12}): $\vec{e} \rightarrow \vec{e} \, \mathrm{sign} \, [eE(t)]$, which brings us to the relation $\Lambda_{\rm lin}=\lambda /2$ instead of Eq.~(\ref{4.14a}). Comparing the right-hand sides of the first equations from the systems (\ref{2.13}) and (\ref{4.1}), we establish the following relation:
\begin{equation}\label{4.15}
U_{\rm lin}=-2 (\vec{e}_{\rm lin} \vec{u}_{\rm lin}).
\end{equation}

Let us return to the general case and introduce the notation
\begin{equation}\label{4.16}
U=-2 (\vec{e} \vec{u}).
\end{equation}
Using Eqs.~(\ref{4.12}) and (\ref{4.16}), one rewrites the first equation of the system~(\ref{4.1}) in the form displayed in the system~(\ref{2.13}):
\begin{equation}\label{4.17}
\dot{f}=\Lambda U.
\end{equation}
In order to examine the remaining equations of the subsystem (\ref{4.1}), we introduce the projection operator with respect to the unit vector $\vec{e}$:
\begin{equation}\label{4.18}
\Delta_{ik}=\delta_{ik}-e_i e_k.
\end{equation}
This operator has the following properties:
\begin{equation}\label{4.19}
\Delta^2=\Delta, \quad \Delta_{ik}e_k=e_i\Delta_{ik}=0, \quad {\rm Sp} \, \Delta = 0.
\end{equation}
Then an arbitrary vector $\vec{a}$ can be represented according to $\vec{a} = a \vec{e} + \vec{a}_\perp$,
%
%\begin{equation}\label{4.20}
%\vec{a} = a \vec{e} + \vec{a}_\perp,
%\end{equation}
where $a=\vec{a} \vec{e}$ and $a_{\perp i}=\Delta_{ik}a_k$. Applying this decomposition to the vectors $\vec{u}$ and $\vec{v}$ from the subsystem (\ref{4.1}) yields the following equations in terms of the longitudinal and transversal components of the vectors $\vec{u}$ and $\vec{v}$:
\begin{equation}\label{4.21}
\begin{aligned}
    \dot{U}+2(\vec{U}_{\perp}\boldsymbol{\cal{E}}) &= 2 \Lambda (1-2f) - 2\omega V, \\
   \dot{V}+2(\vec{V}_{\perp}\boldsymbol{\cal{E}}) &= 2\omega V,
  \end{aligned}
\end{equation}
and
\begin{equation}\label{4.22}
\begin{aligned}
    \dot{\vec{U}}_{\perp} + (\vec{U}_{\perp}\boldsymbol{\cal{E}})\vec{e} + U\boldsymbol{\cal{E}} &= -2 [\vec{U}_{\perp} \boldsymbol{\lambda}_2] - 2\omega \vec{V}_{\perp}, \\
   \dot{\vec{V}}_{\perp} + (\vec{V}_{\perp}\boldsymbol{\cal{E}})\vec{e} + V \boldsymbol{\cal{E}} &= -2 [\vec{V}_{\perp} \boldsymbol{\lambda}_2] + 2\omega \vec{U}_{\perp}.
  \end{aligned}
\end{equation}
In order to keep the analogy with Eq.~(\ref{4.16}), we introduced here the vectors
\begin{equation}\label{4.23}
\vec{U}=-2\vec{u}, \quad \vec{V}=2\vec{v}.
\end{equation}
The vector $\boldsymbol{\cal{E}}$ reads
\begin{equation}\label{4.24}
\boldsymbol{\cal{E}}=\dot{\vec{e}} - 2[\vec{\lambda_2} \vec{e}].
\end{equation}
One can explicitly verify that this vector vanishes,
\begin{equation}\label{4.24a}
\boldsymbol{\cal{E}} = 0~~\Longleftrightarrow~~\dot{\vec{e}}=2[\vec{\lambda_2} \vec{e}],
\end{equation}
once the external field direction is constant,
\begin{equation}\label{4.24b}
\frac{d}{dt}\left(\frac{\vec{E}(t)}{E(t)}\right)=0.
\end{equation}
The linearly polarized field~(\ref{2.12}) is a particular case satisfying the condition~(\ref{4.24b}). The relations~(\ref{4.24a}) turn the system (\ref{4.17}), (\ref{4.21}) into the shortened system~(\ref{2.13}), whereas the system~(\ref{4.22}) becomes homogeneous leading to the trivial solution $\vec{U}_{\perp , \rm lin}=\vec{V}_{\perp , \rm lin}=0$. If the condition~(\ref{4.24a}) holds, the integral of motion~(\ref{4.3}) has the following form (see, e.g, Ref.~\cite{9}):
\begin{equation}\label{4.24c}
(1-2f_{\rm lin})^2+U^2_{\rm lin}+V^2_{\rm lin}=1,
\end{equation}
where we have assumed zero initial conditions.

\subsection{Spin subsystem}

The projection procedure applied to the spin subsystem (\ref{4.2}) brings us to the following equations:
\begin{equation}\label{4.25}
  \begin{aligned}
  \dot{F} - (\vec{F}_{\perp}\boldsymbol{\cal{E}}) &= - 2 \Lambda u, \\
  \dot{\vec{F}}_{\perp} + F \boldsymbol{\cal{E}} + (\vec{F}_{\perp} \boldsymbol{\cal{E}}) &= [\vec{V}_{\perp} \boldsymbol{\lambda}_1] - 2[\vec{F}_{\perp} \boldsymbol{\lambda}_2], \\
  \dot{u} &= 2 \Lambda F + 2 \omega v, \\
  \dot{v} &= -2 \omega u,
  \end{aligned}
\end{equation}
where $\vec{f}=F \vec{e} + \vec{F}_{\perp}$. Since in the case of linear polarization~(\ref{4.24a}), $\vec{V}_{\perp , \rm lin}=0$, the system~(\ref{4.25}) becomes homogeneous, which leads to the trivial solution $F_{\rm lin}=\vec{F}_{\perp , \rm lin}=u_{\rm lin}=v_{\rm lin}=0$, so the spin effects do not take place.

\subsection{Interpretation of the vectors $\boldsymbol{\lambda}_1$ and $\boldsymbol{\lambda}_2$ }

According to Eqs.~(\ref{2.14}) and (\ref{4.14}) the length $\Lambda$ of the vector $\boldsymbol{\lambda}_1$ is the amplitude of the vacuum transitions. The equations~(\ref{4.21}), (\ref{4.22}), and (\ref{4.25}) indicate that the direction of $\boldsymbol{\lambda}_1$ is relevant only to the spin subsystem~(\ref{4.25}), where the function $\boldsymbol{\lambda}_1$ appears together with the coupling function $\vec{V}_{\perp}$.

The vector $\boldsymbol{\lambda}_2$ can be interpreted using the concept of the effective magnetic field~\cite{3a}:
\begin{equation}\label{4.26}
\vec{H}_{\rm eff} = \frac{\gamma}{1 + \gamma}[\vec{E} \vec{v}_\text{g}],
\end{equation}
which takes place in rotating electric fields. Here $\gamma=(1-|\vec{v}_\text{g}|^2)^{-1/2}$. Then, according to Eq.~(\ref{2.8}), $\boldsymbol{\lambda}_2\sim \vec{H}_\text{eff}$ and the vector function $\boldsymbol{\lambda}_2$ describes the rotation of EPP and excitation of the spin degrees of freedom.

\subsection{Reduction to the KE system in graphene}

Using the general KE system (\ref{2.2}), one can derive the KEs for various systems of lower dimensions. As an example, we consider reduction to the KE system regarding the model of low-energy excitations in graphene~\cite{4a}. The simplest model corresponds to (2+1)-dimensional QED dealing with massless fermions (quasielectrons and holes) with the dispersion relation $\omega (\vec{p})=v_\text{F} |\vec{p}|$, where $v_\text{F}$ is the Fermi velocity. A periodical structure of the crystal lattice gives rise to two Dirac points at the boundary of the Brillouin zone. A diagonal form of the Hamiltonian
\begin{equation}\label{4.28}
\hat{H} (t)=v_\text{F} \int \! d^2 x \, \psi^\dagger (\vec{x},t) (\hat{\vec{P}} \boldsymbol{\sigma}) \psi (\vec{x},t)
\end{equation}
can be achieved with the help of a unitary transformation constructed in an explicit form in Ref.~\cite{5a}. This transformation leads to the appearance of anomalous averages and yields the system of three KEs \cite{4a}:
\begin{equation}\label{4.29}
  \begin{aligned}
  \dot{f} &= \frac{1}{2}\lambda_{\rm gr} u, \\
  \dot{u} &= \lambda_{\rm gr} (1-2f)-2\omega_{\rm gr} v, \\
  \dot{v} &= 2\omega_{\rm gr} u,
  \end{aligned}
\end{equation}
where $\omega_{\rm gr}(\vec{p},t) = v_\text{F} |\vec{P}|$. This system has the same form as that displayed in Eqs.~(\ref{2.13}), but the amplitude of vacuum transitions now reads
\begin{equation}\label{4.30}
\lambda_{\rm gr} (\vec{p},t)= \frac{e v_\text{F}^2}{\omega^2_{\rm gr}} [E_1 P_2-E_2 P_1].
\end{equation}
We assume here that the vectors of the electric field and momentum lie in the $xy$ plane,
\begin{equation}\label{4.31}
A^{\mu} = (0,A^1(t),A^2(t),0), \quad \vec{P} = (P^1,P^2,0).
\end{equation}
Note that while the Fermi velocity $v_\text{F}$ now plays the role of the speed of light in the Dirac equation and also enters the dispersion relation, the interaction with the external electric field is introduced by the minimal coupling involving $c$: $P^k=p^k-\frac{e}{c} A^k(t)$, $k=1,2$.

Let us now apply the projection method described in Sect.~\ref{sec:analysis} in order to reduce the general KE subsystem (\ref{4.1}) to the KE system (\ref{4.29}). We assume that the functions $\boldsymbol{\lambda}_1$ and $\boldsymbol{\lambda}_2$ in the system (\ref{4.1}) are defined by introducing the Fermi velocity in Eqs.~(\ref{2.8}) and setting $m=0$. We require then
\begin{equation}\label{4.32}
-2(\boldsymbol{\lambda}_1\vec{u})=\frac{1}{2}\lambda u,~~\text{i.e.}~~u=-(\vec{\Omega}\vec{u}),
\end{equation}
where $\vec{\Omega} \equiv 4\boldsymbol{\lambda}_1/\lambda$. It should be considered as a definition of the as-yet-unknown function $\lambda(\vec{p},t)$. This relation formally ensures the correspondence between the first equations of the systems (\ref{4.1}) and (\ref{4.29}). In accordance with the projection method, we now multiply the second equation of the system (\ref{4.1}) by $-\vec{\Omega}$, which allows one to obtain the second equation in the same form as in (\ref{4.29}), provided
\begin{eqnarray}\label{4.33}
(\vec{\Omega} \boldsymbol{\lambda}_1) &=& \lambda, \\ \label{4.34}%45}
\dot{\vec{\Omega}} - 2[ \boldsymbol{\lambda}_2 \vec{\Omega}] &=& 0.
\end{eqnarray}
The algebraic equation (\ref{4.33}) has the solutions $\lambda=\pm \lambda_{\rm gr}$, 
%
%\begin{equation}\label{4.35}
%\lambda=\pm \lambda_{\rm gr},
%\end{equation}
%
where $\lambda_{\rm gr}$ is defined by Eq.~(\ref{4.30}). Both of the two values provide a CPT invariant structure of the KEs~(\ref{4.29}). One can then straightforwardly verify that the solutions satisfy Eq.~(\ref{4.34}). Finally, the third equation is recovered by setting $v = (\vec{\Omega} \vec{v})$. One can also demonstrate that the vector components of the system~(\ref{4.1}) does not contribute to the KE system in graphene.

\subsection{Energy conservation law} %\label{sec.5}

Following Ref.~\cite{4a}, we will discuss the energy conservation law in 3+1 QED in the case of an arbitrary polarization of the external electric field. If one differentiates the density of quasienergy (\ref{3.19}) with respect to time, one obtains
\begin{equation}\label{4.36}
\dot{E}_\text{q}(t)=\vec{E}(t)\vec{J}(t),
\end{equation}
where $\vec{E}(t)=\vec{E}_{\rm in}(t)+\vec{E}_{\rm ex}(t)$ is the total electric field strength, and the total current $\vec{J}(t)$ is defined in Eqs.~(\ref{3.12})--(\ref{3.13}). Maxwell's equation for the internal (plasma) field reads
\begin{equation}\label{4.37}
\dot{\vec{E}}_{\rm in}(t)=-\vec{J}(t).
\end{equation}
Using Eqs.~(\ref{4.36}) and (\ref{4.37}), one deduces the conservation law
\begin{equation}\label{4.38}
\frac{d}{dt}\left [ E_\text{q}(t) +\frac{1}{2} \vec{E}^2_{\rm in}(t)\right ] = (\vec{E}_{\rm ex} \vec{J}).
\end{equation}
This expression means that the work done by the external field changes both the energy of the quasiparticles and that of the inner field. A specific case of the relation~(\ref{4.38}) was considered in Ref.~\cite{mamaev}. An analogous equation takes place also in graphene~\cite{4a}.

\section{Conclusion} \label{sec.5}

In the present study, we conducted a detailed analysis of the complete system of twelve kinetic equations, which are utilized for describing vacuum production and evolution of EPP under the action of a strong time-dependent spatially homogeneous electric field of arbitrary polarization. First, we presented a brief overview of the well-known properties of the KE system (Sect.~\ref{sec.2}). The physical meaning of all the twelve basic functions was investigated by deriving the macroscopic mean quantities such as current, energy-momentum tensor, and spin tensor (Sect.~\ref{sec.3}). It was also shown that the general KE system represents a combination of two subsystems describing the quasiparticle and spin degrees of freedom, respectively, including the vacuum polarization effects (Sect.~\ref{sec.4}). For these two subsystems, the corresponding integrals of motion were identified. We proposed a projection method which allowed us to systematically derive two particular cases of the KE system: that for the case of a linearly polarized electric field~\cite{5,7,8,9} and for the low-dimensional model of graphene \cite{4a}. Finally, in Appendix an instantaneous basis approach was described in detail and employed in order to deduce the general KE system.

This investigation reveals more detailed information on the properties of the KE system in the case of general time-dependent external fields and should enhance our understanding of how the 12 basic components are related to the physical characteristics of EPP.

%This investigation was motivated by the attempts to correctly describe the spin effects in a strong circularly polarized electric field~\cite{6a}, by the problem of how the kinetic approach is related to the imaginary time method~\cite{7a}, and also by the need for a description of $e^+e^-\gamma$ plasma in the case of electric backgrounds of arbitrary polarization.

\begin{acknowledgement}

S.A.S. and V.V.D. wish to express their gratitude to D.~B.~Blaschke on the occasion of his birthday anniversary. We thank him for the longstanding and fruitful collaboration and support. His diverse interests and energy have had a positive and profound influence on the development of kinetic methods with regard to QFT phenomena in strong fields.

S.A.S. also thanks A.~M.~Fedotov for valuable discussions of some problems regarding the present study. The work of V.V.D. and S.A.S. was supported by the RFBR research project~No.~18-07-00778. I.A.A. acknowledges the support from the Foundation for the advancement of theoretical physics and mathematics ``BASIS''.

\end{acknowledgement}

\appendix

\section*{Appendix: Instantaneous basis approach}

The main idea of this approach is to decompose the field operator by means of the basis set of the instantaneous eigenfunctions of the one-particle Hamiltonian $\mathcal{H}(t) = \boldsymbol{\alpha}[-i \vec{\nabla} - e\vec{A}(t)] + \beta m$ and formulate the problem in terms of the corresponding coefficients. They contain all the information about the production probabilities, and the system of differential equations involving these coefficients is equivalent to the system~(\ref{2.2}).

We introduce the orthonormal and complete set of the eigenfunctions of the Hamiltonian at each given time instant $t$, so these functions have not only spatial dependence, which is trivial due to the homogeneity of the external field, but also implicit temporal dependence since the Hamiltonian incorporates a nonstatic external field. The explicit form of the functions of the instantaneous basis reads
\begin{equation} \tag{A.1}
  \begin{aligned}
    \phi^{+}_{\vec{p},s}(\vec{x},t) &= (2\pi)^{-3/2}\,
    \mathrm{e}^{i\vec{p} \vec{x}}
    u_{s}(\vec{p},t),
    \\
    \phi^{-}_{\vec{p},s}(\vec{x},t) &= (2\pi)^{-3/2}\,
    \mathrm{e}^{-i\vec{p} \vec{x}}
    v_{s}(\vec{p},t),
  \end{aligned}
  \qquad
    \begin{aligned}
    {}^{+}\epsilon(\vec{p},t) &= \omega(\vec{p},t),
    \\
    {}^{-}\epsilon(\vec{p},t) &= -\omega(-\vec{p},t),
  \end{aligned}
\label{inst_func_explicit}
\end{equation}
where we will employ the bispinors from Eqs.~(\ref{2.9}), and the plus (minus) sign indicates that the corresponding energy ${}^{\pm}\epsilon(\vec{p},t)$ is positive (negative). Let us define two sets of solutions of the time-dependent Dirac equation $\{{}_{\pm}\Phi_{\vec{p},s}(x)\}$ and $\{{}^{\pm}\Phi_{\vec{p},s}(x)\}$ determined by the following conditions at $t=t_{\text{in}}$ and $t=t_{\text{out}}$, when the external field is switched on and off, respectively:
\begin{equation}\tag{A.2}
  {}_{\pm}\Phi_{\vec{p},s}(\vec{x},t_{\text{in}}) =
  \phi^{+}_{\vec{p},s}(\vec{x},t_{\text{in}}),\quad
  {}^{\pm}\Phi_{\vec{p},s}(\vec{x},t_{\text{out}}) =
  \phi^{-}_{\vec{p},s}(\vec{x},t_{\text{out}}).
\end{equation}
In what follows, we will refer to these two sets as the \textit{in} and \textit{out} solutions, respectively. They can be used to decompose the field operator. For instance, in terms of the {\it in} solutions, the decomposition reads
\begin{equation}
  \psi(\vec{x}, t) = \sum\limits_s \! \int \! d^3p \bigl[
  a_{\text{in},s}(\vec{p})\,{}_+\Phi_{\vec{p},s}(\vec{x}, t)+
  b^\dagger_{\text{in},s}(\vec{p})\,{}_-\Phi_{\vec{p},s}(\vec{x}, t) \bigr],\tag{A.3}
  \label{decomp_field_in}
\end{equation}
which corresponds to the Heisenberg representation. In terms of the instantaneous basis, the field operator is expanded as
\begin{equation}
  \psi(\vec{x}, t) = \sum\limits_s \! \int \! d^3p \bigl[
  a_{s}(\vec{p},t)\phi^{+}_{\vec{p},s}(\vec{x}, t)+
  b^\dagger_s(\vec{p},t)\phi^{-}_{\vec{p},s}( \vec{x}, t)\bigr].\tag{A.4}
  \label{eq:decomposition_inst}
\end{equation}
This expansion immediately leads to the representation of the field Hamiltonian displayed in Eq.~(\ref{2.11}).

One can demonstrate that the number density of electrons produced with momentum $\vec{p}$ in the spin state $s$ can be evaluated via
\begin{equation}\label{number_of_particles}\tag{A.5}
  \frac{dN_{\vec{p}, s}}{d^3p} = \lim_{t\rightarrow t_{\text{out}}}\langle
  a^{\dagger}_{s}(\vec{p},t) a_{s}(\vec{p},t) \rangle,
\end{equation}
where the averaging corresponds to the \textit{in} vacuum state defined by means of the \textit{in} creation/annihilation operators which appear in Eq.~(\ref{decomp_field_in}). In order to calculate the vacuum expectation value in Eq.~(\ref{number_of_particles}), one should express the time-dependent operators $a_{s}(\vec{p},t)$ in terms of the \textit{in} operators. To this end, we decompose the \textit{in} solutions with the aid of the instantaneous basis:
\begin{align}\label{decomp_inout_1}\tag{A.6}
  {}_+\Phi_{\vec{p},s}(\vec{x}, t) &= \sum_{s'} \int \! d^3q \left[
  \mathfrak{d}_{ss'}(\vec{p},\vec{q},t) \phi^{+}_{\vec{q},s'}(\vec{x},t) +
  \mathfrak{e}_{ss'}(\vec{p},\vec{q},t) \phi^{-}_{\vec{q},s'}(\vec{x},t) \right],
  \\ \label{decomp_inout_2}\tag{A.7}
  {}_-\Phi_{\vec{p},s}(\vec{x}, t) &= \sum_{s'} \int \! d^3q \left[
  \mathfrak{f}_{ss'}(\vec{p},\vec{q},t) \phi^{+}_{\vec{q},s'}(\vec{x},t) +
  \mathfrak{g}_{ss'}(\vec{p},\vec{q},t) \phi^{-}_{\vec{q},s'}(\vec{x},t) \right].
\end{align}
%
%The corresponding coefficients read
%%
%\begin{align}\tag{A.9}
%  \mathfrak{d}_{ss'}(\vec{p},\vec{q},t) &= \bigl(
%  \phi^{+}_{\vec{q},s'},\,{}_+\Phi_{\vec{p},s}\bigr),
%  \quad
%  \mathfrak{e}_{ss'}(\vec{p},\vec{q},t) = \bigl(
%  \phi^{-}_{\vec{q},s'},\,{}_+\Phi_{\vec{p},s}\bigr),
%  \\ \tag{A.10}
%  \mathfrak{f}_{ss'}(\vec{p},\vec{q},t) &= \bigl(
%  \phi^{+}_{\vec{q},s'},\,{}_-\Phi_{\vec{p},s}\bigr),
%  \quad
%  \mathfrak{g}_{ss'}(\vec{p},\vec{q},t) = \bigl(
%  \phi^{-}_{\vec{q},s'},\,{}_-\Phi_{\vec{p},s}\bigr),
%\end{align}
%%
%where we imply the ordinary inner product $(\psi_{1},\psi_{2}) = \displaystyle{\int} d^3 x \, \psi^{\dagger}_{1} \psi_{2}$.
%
Since the external field does not depend on the coordinates, the spatial dependence of the \textit{in} solutions is also evident:
\begin{equation}\tag{A.8}
  {}_\pm\Phi_{\vec{p},s}(\vec{x}, t) =
  (2\pi)^{-3/2} \mathrm{e}^{\pm i \vec{p} \vec{x}} \, {}_\pm\Phi_{s}(\vec{p},t).
\end{equation}
As a result, the coefficients have a diagonal form with respect to momentum:
\begin{align}\tag{A.9}
  \mathfrak{d}_{ss'}(\vec{p},\vec{q},t) &=
  \delta(\vec{p}-\vec{q}) u^{\dagger}_{s'}(\vec{p},t) {}_+\Phi_s(\vec{p},t) \equiv \delta(\vec{p}-\vec{q})
  \mathfrak{d}_{ss'}(\vec{p},t) ,
  \\ \tag{A.10}
  \mathfrak{e}_{ss'}(\vec{p},-\vec{q},t) &=
  \delta(\vec{p}-\vec{q}) v^{\dagger}_{s'}(-\vec{p},t) {}_+\Phi_s(\vec{p},t) \equiv \delta(\vec{p}-\vec{q}) \mathfrak{e}_{ss'}(-\vec{p},t),
  \\ \tag{A.11}
  \mathfrak{f}_{ss'}(\vec{p},-\vec{q},t) &=
  \delta(\vec{p}-\vec{q}) u^{\dagger}_{s'}(-\vec{p},t) {}_-\Phi_s(\vec{p},t) \equiv \delta(\vec{p}-\vec{q})
  \mathfrak{f}_{ss'}(-\vec{p},t),
  \\ \tag{A.12}
  \mathfrak{g}_{ss'}(\vec{p},\vec{q},t) &=
  \delta(\vec{p}-\vec{q}) v^{\dagger}_{s'}(\vec{p},t) {}_-\Phi_s (\vec{p},t) \equiv \delta(\vec{p}-\vec{q})
  \mathfrak{g}_{ss'}(\vec{p},t).
\end{align}
Using these expressions and plugging then the decompositions~(\ref{decomp_inout_1}) and (\ref{decomp_inout_2}) into Eq.~(\ref{decomp_field_in}), one establishes the following relations:
%
%Thus, Eqs.~(\ref{decomp_inout_1}) and (\ref{decomp_inout_2}) take the following form:
%
%\begin{align}\tag{A.16}
%  {}_+\Phi_{\vec{p},s}(\vec{x}, t) &= \sum_{s'} \left[
%  \mathfrak{d}_{ss'}(\vec{p},t) \phi^{+}_{\vec{p},s'}(\vec{x},t) +
%  \mathfrak{e}_{ss'}(-\vec{p},t) \phi^{-}_{-\vec{p},s'}(\vec{x},t) \right], \label{Phi_in_inst_1}
%  \\ \tag{A.17}
%  {}_-\Phi_{\vec{p},s}(\vec{x}, t) &= \sum_{s'} \left[
%  \mathfrak{f}_{ss'}(-\vec{p},t) \phi^{+}_{-\vec{p},s'}(\vec{x},t) +
%  \mathfrak{g}_{ss'}(\vec{p},t) \phi^{-}_{\vec{p},s'}(\vec{x},t) \right]. \label{Phi_in_inst_2}
%\end{align}
%
%Plugging these decompositions into Eq.~(\ref{decomp_field_in}) and comparing the resulting expression with Eq.~(\ref{eq:decomposition_inst}), one establishes the following relations:
%
\begin{align}\tag{A.13}
  a_{s}(\vec{p},t) &= \sum_{s'}\left[ \mathfrak{d}_{s's}(\vec{p},t)
  a_{\text{in},s'}(\vec{p}) +
  \mathfrak{f}_{s's}(\vec{p},t) b^{\dagger}_{\text{in},s'}(-\vec{p}) \right], \label{inst_in_a}
  \\ \tag{A.14}
  b^{+}_{s}(\vec{p},t) &= \sum_{s'}\left[ \mathfrak{e}_{s's}(\vec{p},t)
  a_{\text{in},s'}(-\vec{p})
  +
  \mathfrak{g}_{s's}(\vec{p},t) b^{\dagger}_{\text{in},s'}(\vec{p}) \right]. \label{inst_in_b}
\end{align}
Using then Eqs.~(\ref{number_of_particles}) and (\ref{inst_in_a}), one obtains the following expression for the number density of particles created:
\begin{equation} \tag{A.15}
  \frac{(2\pi)^3}{V} \frac{dN_{\vec{p}, s}}{d^3p} =
  \lim_{t\rightarrow t_{\text{out}}}
  \sum_{s'} |\mathfrak{f}_{s's}(\vec{p},t)|^2.
\end{equation}
Our aim is to derive equations which govern the evolution of the function $\mathfrak{f}_{s,s'}$. It can be achieved if one makes use of the fact that the function ${}_-\Phi_{\vec{p},s}$ is a solution of the Dirac equation:
\begin{equation}\label{dirac_operator}\tag{A.16}
  [i\partial_{t} - \mathcal{H}(t)]\sum_{s'}
  \left[
  \mathfrak{f}_{ss'}(\vec{p},t) \phi^{+}_{\vec{p},s'}(\vec{x},t) +
  \mathfrak{g}_{ss'}(-\vec{p},t) \phi^{-}_{-\vec{p},s'}(\vec{x},t)
  \right] = 0.
\end{equation}
Here we have changed the sign of $\vec{p}$. Using then the explicit form~(\ref{inst_func_explicit}) of the instantaneous eigenfunctions and projecting Eq.~(\ref{dirac_operator}) onto this basis set, we receive the following system of equations:
\begin{equation} \label{fg_system_bispinors}\tag{A.17}
\begin{aligned}
  i\dot{\mathfrak{f}}_{s,1} +
  i\mathfrak{f}_{s,1} u^{\dagger}_{1} \dot{u}_{1} +
  i\mathfrak{f}_{s,2} u^{\dagger}_{1} \dot{u}_{2} +
  i\mathfrak{g}_{s,1} u^{\dagger}_{1} \dot{v}_{1} +
  i\mathfrak{g}_{s,2} u^{\dagger}_{1} \dot{v}_{2}
  &= \omega(\vec{p}) \mathfrak{f}_{s,1},
  \\
  i\dot{\mathfrak{f}}_{s,2} +
  i\mathfrak{f}_{s,1}u^{\dagger}_{2} \dot{u}_{1} +
  i\mathfrak{f}_{s,2} u^{\dagger}_{2} \dot{u}_{2} +
  i\mathfrak{g}_{s,1} u^{\dagger}_{2} \dot{v}_{1} +
  i\mathfrak{g}_{s,2} u^{\dagger}_{2} \dot{v}_{2}
  &= \omega(\vec{p}) \mathfrak{f}_{s,2},
  \\
  i\dot{\mathfrak{g}}_{s,1} +
  i\mathfrak{f}_{s,1} v^{\dagger}_{1} \dot{u}_{1} +
  i\mathfrak{f}_{s,2} v^{\dagger}_{1} \dot{u}_{2} +
  i\mathfrak{g}_{s,1} v^{\dagger}_{1} \dot{v}_{1} +
  i\mathfrak{g}_{s,2} v^{\dagger}_{1} \dot{v}_{2}
  &= -\omega(\vec{p}) \mathfrak{g}_{s,1},
  \\
  i\dot{\mathfrak{g}}_{s,2} +
  i\mathfrak{f}_{s,1} v^{\dagger}_{2} \dot{u}_{1} +
  i\mathfrak{f}_{s,2} v^{\dagger}_{2} \dot{u}_{2} +
  i\mathfrak{g}_{s,1} v^{\dagger}_{2} \dot{v}_{1} +
  i\mathfrak{g}_{s,2} v^{\dagger}_{2} \dot{v}_{2}
  &= -\omega(\vec{p}) \mathfrak{g}_{s,2},
\end{aligned}
\end{equation}
where we have omitted the arguments of the time-dependent functions and of the bispinors. The value of $s$ determines the initial conditions, $\mathfrak{g}_{ss'}(-\vec{p},t_{\text{in}}) = \delta_{ss'}$. We combine then the $s'$ components defining the vector $\mathfrak{f}^{\dagger}_{s} = [\mathfrak{f}_{s,1}^{*}, \mathfrak{f}_{s,2}^{*}]$ (similar expression is introduced for~$\mathfrak{g}^{\dagger}_{s}$). The system~(\ref{fg_system_bispinors}) can be now rewritten using matrix notation:
\begin{equation} \label{fg_system_M}\tag{A.18}
\begin{aligned}
  \dot{\mathfrak{f}}_{s}(\vec{p},t) &=
  M_{1}\mathfrak{f}_{s}(\vec{p},t) + M_{2}\mathfrak{g}_{s}(-\vec{p},t),
  \\
  \dot{\mathfrak{g}}_{s}(-\vec{p},t) &=
  M_{3}\mathfrak{f}_{s}(\vec{p},t) + M_{4}\mathfrak{g}_{s}(-\vec{p},t).
\end{aligned}
\end{equation}
The form of the matrices $M_i$ will be discussed below. Although this system can already be utilized in numerical computations, we will formulate the equations in terms of the following $2\times 2$ matrices:
%
%\begin{equation}%\tag{A24}
%  f_{ss'}(\vec{p},t) = \frac{(2\pi)^3}{V} \, \langle
%  a^{\dagger}_{s'}(\vec{p},t) a_{s}(\vec{p},t) \rangle_{\text{in}}, \quad
%  g_{ss'}(\vec{p},t) = \frac{(2\pi)^3}{V} \, \langle
%  b_{s'}(-\vec{p},t) b^{\dagger}_{s}(-\vec{p},t) \rangle_{\text{in}},
%\end{equation}
%\begin{equation}%\tag{A25}
%  y^{+}_{ss'}(\vec{p},t) = \frac{(2\pi)^3}{V} \,\langle
%  a^{\dagger}_{s'}(\vec{p},t) b^{\dagger}_{s}(-\vec{p},t) \rangle_{\text{in}},
%  \quad
%  y^{-}_{ss'}(\vec{p},t) = \frac{(2\pi)^3}{V} \,\langle
%  b_{s'}(-\vec{p},t) a_{s}(\vec{p},t) \rangle_{\text{in}}.
%\end{equation}
%%
%Using the spinor functions $\mathfrak{f}_{s}$ and $\mathfrak{g}_{s}$, one can represent these relations in the form
%
\begin{equation}\tag{A.19}
  \begin{aligned}
    f(\vec{p},t) &= \sum_{s}
    \mathfrak{f}_{s}(\vec{p},t) \mathfrak{f}^{\dagger}_{s}(\vec{p},t),\quad
    g(\vec{p},t) = \sum_{s}
    \mathfrak{g}_{s}(-\vec{p},t) \mathfrak{g}^{\dagger}_{s}(-\vec{p},t),
    \\
    y^{+}(\vec{p},t) &= \sum_{s}
    \mathfrak{g}_{s}(-\vec{p},t) \mathfrak{f}^{\dagger}_{s}(\vec{p},t),\quad
    y^{-}(\vec{p},t) = \sum_{s}
    \mathfrak{f}_{s}(\vec{p},t) \mathfrak{g}^{\dagger}_{s}(-\vec{p},t).
  \end{aligned}
\end{equation}
The system~(\ref{fg_system_M}) now reads
\begin{equation}\tag{A.20}
  \begin{aligned}
    \dot{f}(\vec{p},t) &= M_{1}f(\vec{p},t) + f(\vec{p},t)M_{1}^{\dagger} +
    M_{2}y^{+}(\vec{p},t) + y^{-}(\vec{p},t)M_{2}^{\dagger},
    \\
    \dot{g}(\vec{p},t) &= M_{4}g(\vec{p},t) + g(\vec{p},t)M_{4}^{\dagger} +
    M_{3}y^{-}(\vec{p},t) + y^{+} (\vec{p},t) M_{3}^{\dagger},
    \\
    \dot{y}^{+}(\vec{p},t) &=  M_{4}y^{+}(\vec{p},t) +
    y^{+}(\vec{p},t)M_{1}^{\dagger} +
    M_{3}f(\vec{p},t) + g(\vec{p},t)M_{2}^{\dagger},
    \\
    \dot{y}^{-}(\vec{p},t) &= M_{1}y^{-}(\vec{p},t) +
    y^{-}(\vec{p},t)M_{4}^{\dagger} +
    M_{2}g(\vec{p},t) + f(\vec{p},t)M_{3}^{\dagger}.
  \end{aligned}
\end{equation}
Instead of 8 complex components, we have now 16 which possess, however, symmetry properties regarding Hermitian conjugation, i.e., the matrices $f(\vec{p},t)$ and $g(\vec{p},t)$ are Hermitian and $y^{-}(\vec{p},t)$ is the Hermitian transpose of $y^{+}(\vec{p},t)$. Therefore, the number of the independent components remains the same. One can explicitly verify the following relations:
\begin{align}\tag{A.21}
  M_{1} &= M^{\dagger} - i\omega \mathbb{I}, \quad M_{4} = M^{\dagger} + i\omega \mathbb{I},
  \\ \tag{A.22}
  M_{3} &= -M_{2} = \frac{e}{2\omega^{2}\omega_{+}}\big [
  \vec{P}(\vec{P}\vec{E}) - \omega\omega_{+} \vec{E} \big ]\boldsymbol{\sigma} \equiv \Xi.
\end{align}
Moreover, the matrix elements of $M$ obey $M_{11} = M^*_{22}$ and $M_{21} = -M^*_{12}$. The real part of the diagonal elements of this matrix vanishes and the residual part of $M$ becomes antihermitian. Taking this into account, we arrive at
%
%\begin{equation}\label{system}%\tag{A31}
%  \begin{aligned}
%    \dot{f}(\vec{p},t) &= f(\vec{p},t) M + M^{\dagger} f(\vec{p},t) -
%    \left[ \Xi y^{+}(\vec{p},t) + y^{-}(\vec{p},t) \Xi \right],
%    \\
%    \dot{g}(\vec{p},t) &= g(\vec{p},t) M + M^{\dagger} g(\vec{p},t) +
%    \left[ y^{+}(\vec{p},t) \Xi + \Xi y^{-}(\vec{p},t) \right],
%    \\
%    \dot{y}^{+}(\vec{p},t) &= y^{+}(\vec{p},t) M + M^{\dagger} y^{+}(\vec{p},t) +
%    \left[ \Xi f(\vec{p},t) -g(\vec{p},t) \Xi \right]+2i\omega y^{+}(\vec{p},t),
%    \\
%    \dot{y}^{-}(\vec{p},t) &= y^{-}(\vec{p},t) M + M^{\dagger} y^{-}(\vec{p},t) +
%    \left[ f(\vec{p},t)\Xi -\Xi g(\vec{p},t) \right]
%    -2i\omega y^{-}(\vec{p},t).
%  \end{aligned}
%\end{equation}
%
%One can also show that the matrix elements of $M$ obey $M_{11} = M^*_{22}$ and
%$M_{21} = -M^*_{12}$. The real part of the diagonal elements of this matrix
%vanishes and the residual part of $M$ becomes antihermitian. Finally, we obtain
%
\begin{equation}\label{system_fgy_final}\tag{A.23}
  \begin{aligned}
    \dot{f}(\vec{p},t) &= [f(\vec{p},t),\Theta] -
    \left[ \Xi y^{+}(\vec{p},t) + y^{-}(\vec{p},t) \Xi \right],
    \\
    \dot{g}(\vec{p},t) &= [g(\vec{p},t),\Theta] +
    \left[ y^{+}(\vec{p},t) \Xi + \Xi y^{-}(\vec{p},t) \right],
    \\
    \dot{y}^{+}(\vec{p},t) &= [y^{+}(\vec{p},t),\Theta] +
    \left[ \Xi f(\vec{p},t) -g(\vec{p},t) \Xi \right]
    +2i\omega y^{+}(\vec{p},t),
    \\
    \dot{y}^{-}(\vec{p},t) &= [y^{-}(\vec{p},t),\Theta] +
    \left[ f(\vec{p},t)\Xi -\Xi g(\vec{p},t) \right]
    -2i\omega y^{-}(\vec{p},t),
  \end{aligned}
\end{equation}
where
\begin{equation}\tag{A.24}
  \Theta \equiv \frac{ie [\vec{P} \vec{E}] }
  {2\omega\omega_{+}} \, \boldsymbol{\sigma}.
\end{equation}
At time instant $t = t_{\text{in}}$, we have $g(\vec{p},t_{\text{in}}) = \mathbb{I}$ while all of the other matrices vanish. The system~(\ref{system_fgy_final}) was first derived in Ref.~\cite{11}. If one expresses the matrices in the Pauli representation, this system will turn to that presented in Eq.~(\ref{2.2}). This transformation is described in detail, e.g., in Ref.~\cite{14}.

\end{document}